\documentstyle[epsfig]{elsart}

\makeatletter
\if@twoside
\else \oddsidemargin 00pt \evensidemargin 00pt
\fi
\let\@eqnsel = \hfil

\expandafter\ifx\csname mathrm\endcsname\relax\def\mathrm#1{{\rm #1}}\fi
\@ifundefined{mathrm}{\def\mathrm#1{{\rm #1}}}{\relax}

\makeatother

\begin{document}

\begin{frontmatter}
\title {{\bf  Single chargino production with  
R-parity lepton number violation
in electron-electron and muon-muon  collisions}}

\author[ad1]{Micha{\l}  Czakon}
and
\author[ad1,ad2]{Janusz Gluza}
\address[ad1]{ Department of Field Theory and Particle Physics, University of
Silesia, Uniwersytecka 4, PL-40-007 Katowice, Poland}
\address[ad2]{ Deutsches Elektronen-Synchrotron DESY, Zeuthen,  Germany}
\begin{abstract}
We examine  single chargino production in conjunction with R-parity lepton
number violation in future lepton-lepton collisions. 
Present bounds on R-parity violating couplings allow 
for a production cross section of the order of ${\cal{O}} ( 10 \mbox{ fb})$
for a wide range of sneutrino and chargino masses.
Scenarios of chargino decay which lead to purely leptonic signals in
the final state and without missing energy are also discussed.
\end{abstract}
\end{frontmatter}

\section{Introduction}
R-parity is a discrete
symmetry defined by assigning to every field the number $R=(-1)^{3B+L+2S}$ ($B (L)$ - baryon (lepton)
number, $S$ - spin of the particle) \cite{far}.
If it is conserved then baryon and lepton number violating 
transitions are forbidden. In that case, the theory guarantees both 
proton stability and lepton universality. 
However, in supersymmetric extensions of the Standard Model, gauge invariance 
and renormalizability, the two main principles 
of any gauge theory, do not assure R-parity conservation.
At present, wide phenomenological investigations of  R-parity violating 
processes have been undertaken  (for reviews see e.g.  \cite{dr,rep}).

Here we will explore the possibility of discovering the lepton number violating 
process of single chargino production
at future lepton-lepton colliders (see Fig.1(I) for an electron-electron 
collision). 
To our knowledge this process  has not yet  been  discussed,
though lepton number violating charginos pair production 
in electron-electron collisions (Fig.1(II)) has been considered 
\cite{cu,th,hi1}.
\begin{figure}[h]
\epsfig{figure=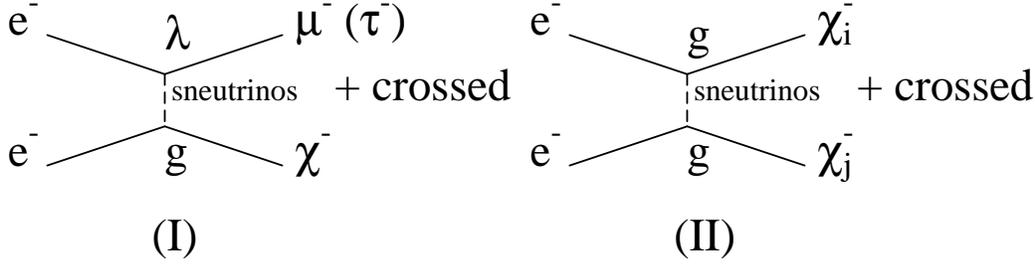, height=1.5 in}
\caption{Single (I) and double (II) chargino production at an $e^-e^-$ 
collider. $\lambda$,g are couplings specified in Eqs. (1,2).}
\end{figure}

Let us start with electron-electron collisions. The analysis of the muon option
is analogous and will be shortly discussed whenever needed.
As can be  seen from Fig.1(I), the cross section for single chargino 
production is proportional to $\lambda^2g^2$ where $\lambda$ and $g$ are 
couplings involved in  the
following Lagrangians ($\lambda_{abc}=-\lambda_{acb}$, $a,b,c$ are 
family indices):
\begin{eqnarray}
{\cal L}&=&
{g}\bar{\tilde{\chi}}^c_i V_{i1} 
(1-\gamma_5)e\tilde{K}_{em}\tilde{\nu}_{m}^{\ast}
+ h.c. ,
\label{lag_g}  \\
&& \nonumber \\
 {\cal{L}}_{\not R_p} &=&
\lambda_{abc} \{\tilde \nu_{aL} \bar l_{cR} l_{bL} 
-(a \leftrightarrow b )\} + h.c.
\label{lam}
\end{eqnarray}
These Lagrangians are written in physical basis.
The matrix $\tilde{K}_{em}$ in Eq.~(\ref{lag_g})
comes from the sneutrino mass matrix diagonalization.
If R-parity is violated,  we have
to take into account the 
mixing between the sneutrinos $\tilde{\nu}_e$, $\tilde{\nu}_{\mu}$, 
$\tilde{\nu}_{\tau}$ and the neutral Higgs bosons $H^0_1$, $H^0_2$.
We shall, however, assume that this mixing
is negligible and does not affect the results,
at least at the stage of chargino production.
In what follows we shall  also assume that the 
exchange of the lightest (electron) sneutrino dominates 
(which is equivalent to 
some hierarchy assumption in the sneutrino sector) and neglect the 
contribution of the heavier $\tilde{\nu}$'s. We therefore set 
($e$ stands for electron)
$\tilde{K}_{em}=\delta_{em}$ in Eq.~(\ref{lag_g}).
For more complicated cases where the interplay between sneutrino masses in
propagators and appropriate elements of the $\tilde{K}$ matrix matters 
we refer to \cite{th}.
 
The second mixing matrix, namely  $V_{i 1}$ in Eq.~(\ref{lag_g}) 
is connected with the chargino sector and describes the weights of
the wino component of the chargino fields \cite{hans}. 
Since this is the only  component
of the charginos that couples to the electron and the sneutrino
(the charged higgsino coupling is
neglected in the limit of zero electron mass) we set for simplicity
$V_{i 1}=1$. 
This is further justified by the analysis \cite{fut} (in the parameter region 
$|\mu | \geq 100$ GeV, $M_2 \geq 100$ GeV for  both small and large
$\tan \beta $, with $\mu, M_2$ being the 
higgsino and gaugino  $SU(2)$ mass parameters, respectively, and $\beta$
a ratio of two vacuum expectation values involved in MSSM).
In general the results should be multiplied by $V_{i 1}^2$.
Furthermore, with  R-parity violation, 
additional couplings between  leptons, gauginos and higgsinos 
$( e, \mu , \tau , \tilde W^- , \tilde H^- )_L$  exist, but  are known 
to be smaller than the gauge ones
\cite{bart}.

\section{Single chargino production and decays: results}

\vspace{.5 cm} 
\begin{figure}[h]
\epsfig{figure=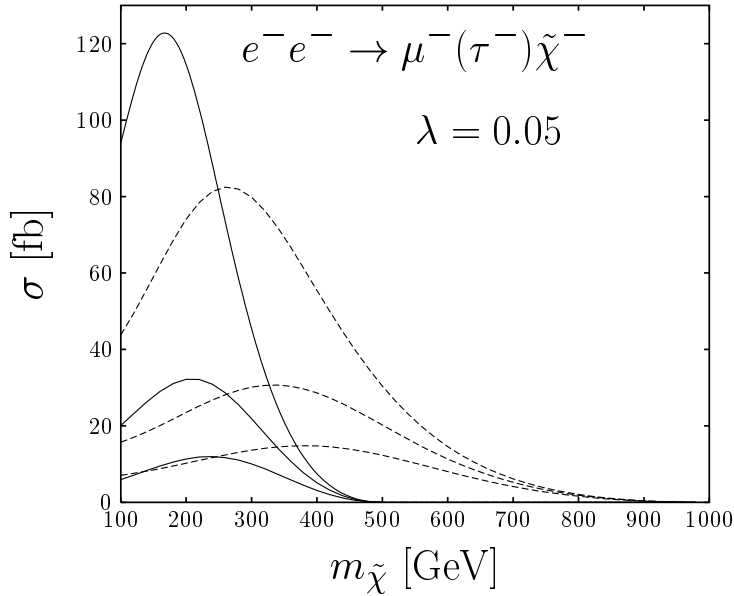, height=3. in}
\caption{Production of a single chargino in $e^-e^- \rightarrow \mu^-(\tau^-)
 \tilde
\chi^-$  as function of its mass for both $\sqrt{s}=500$ GeV
(solid) and $\sqrt{s}=1$ TeV (dashed) energies, $\lambda_{112(3)}=0.05$. 
In both cases, curves  corresponding to sneutrino masses 
$m_{{\tilde \nu}_e}=100,200,300$ GeV are given. A 100\% left-handed electron 
beam is assumed and the chargino is a pure wino state (see discussion in the 
introduction).} 
\end{figure}

In Fig. 2 we gather the cross sections for single chargino production
at future electron-electron colliders with c.m. energies 
$\sqrt{s}=500$ GeV and $\sqrt{s}=1$ TeV
as functions of the chargino mass for different sneutrino
masses\footnote{Results (see Appendix) 
assume a $P_-^e=-100\%$ electron beam polarization. 
In reality we
can expect that $P_-^e=-90\%$  can be achieved. Then the cross sections
 must be multiplied  by a factor $\frac{1}{4}(1-P_-^{e_1})(1-P_-^{e_2}) \simeq
0.9$.}.
For the R-parity violating coupling, we have used  the most conservative
available upper limit $\lambda_{112(3)} \equiv \lambda=0.05$ \cite{str}, 
independently of the $\tilde{\nu}_e$ mass (in the case of muon-muon collisions the
$\lambda_{212(3)}$ couplings would be involved).
For sneutrino masses larger than 100 GeV this limit becomes weaker \cite{str}. 

As can be deduced from Fig.2, 
with a planned annual luminosity of some 50 fb$^{-1}$ yr$^{-1}$ \cite{ecfa}
 and with a discovery
limit at a level of 10 events per year ($\sigma = 0.2$ fb),
the process is detectable 
 for a wide range of sparticle masses.

With the R-parity violating 
production process (I) we are already definitely out of the
SM physics. 
It is therefore interesting to investigate the possible detector signals.
With R-parity non-conservation, the
collider phenomenology is quite different from the MSSM case and depends
especially 
on the nature of the LSP (Lightest Supersymmetric Particle). In the MSSM, 
the stable LSP must be charge and 
color neutral  for cosmological reasons \cite{el}. With R-parity 
violation there are no hints about the unstable LSP. 
It can be among others a sneutrino, gluino or even a chargino \cite{dr,kolb}.
\begin{figure}[h]
\epsfig{figure=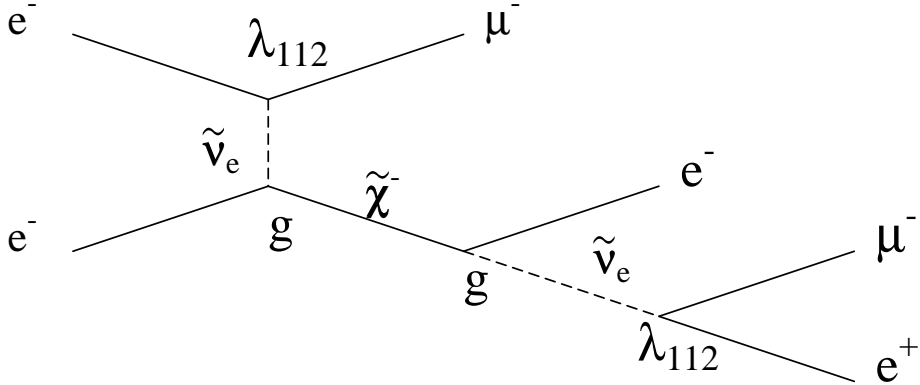, height=2.2 in}
\caption{Possible signature for lepton number violation which is discussed in
the text. For appropriate  numerics see Fig.4.}
\end{figure}
Here we give an example of nonstandard phenomenology but
restrict ourselves to a scenario   
in which charginos decay uniquely (via sneutrino exchange) 
to charged leptons. Final leptonic signals with 
lepton number violation and without missing energy could be detected,
an interesting situation  from the point of view of
nonstandard physics,  as
there is no  SM background  (see further discussion).
These two conditions (charged leptons without missing 
energy in the final state) require the chargino to be the second 
lightest supersymmetric particle
(NLSP) with sneutrino the LSP. This situation is schematically summarized in 
Fig.3. If the chargino were the LSP its lifetime should be long enough so that
it would be seen in the detector.
In other cases 
(i.e. when the chargino is neither  NLSP nor LSP) 
the  chargino would also have cascade decays to  final jet states  
\cite{lol}. Then, the situation would be more complicated
but at least we can expect that for kinematical reasons
a decay to the R-parity lepton violating LSP sneutrino 
would still be important  and  the final 
signal with four charged  leptons could be observed (work in progress).
 
Let us discuss the scenario with NLSP chargino.
First, we should notice that,  to get substantial chargino 
production (e.g. $e^-e^- \to \mu^- \tilde{\chi}^-$ in Fig.2),
 we are interested in the situation where at least 
one, let us say  $\lambda_{112}$ coupling is large.
Then the decay of chargino
to three charged leptons must be observed in the detector, 
as it can undergo uniquely 
through the same large $\lambda_{112}$ coupling (the only possible decay
channel, see  Fig.3).


\vspace{.5 cm}
\begin{figure}[h]
\epsfig{figure=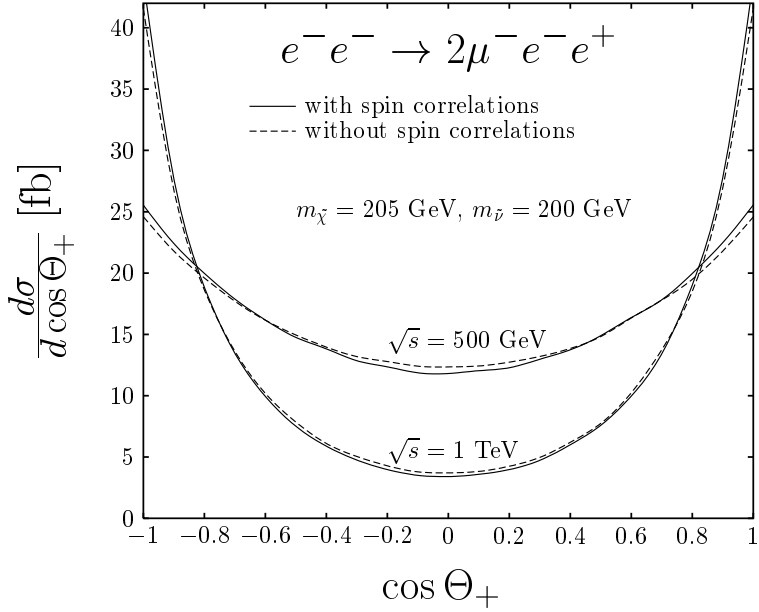, height=3.1 in}
\caption{Angular distribution of the final positron from Fig.3 with  
spin correlations taken into account (solid line) and with factorization 
 between its production and decay (dashed line).
$\sqrt{s}=1$ TeV, $m_{\tilde \chi}=205$ GeV, $m_{\tilde \nu}=200$ GeV,
$\lambda=0.05$.}
\end{figure}
In Fig.4, we show the final results for the  angular distribution of the final 
positron (Fig.3) for two different energies ($\sqrt{s}=500(1000)$ GeV). 
We have taken $m_{\tilde \chi}=205$ GeV and 
$m_{\tilde \nu}=200$ GeV.
Results have been obtained using the VEGAS 
procedure. Four particles in the final state give us an 8 dimensional 
integration. 
We have also applied the  narrow width approximation where
$\Gamma_{\tilde \chi} <<
m_{\tilde \chi}$ (see Appendix for details). 
The solid line describes results based on Eq. (A.21), 
when interferences between
production and decay of charginos with $\tilde \lambda=\pm 1/2$ are taken into
account. The dashed line describes results with factorization  assumed 
\cite{fact},
which means the following replacement in Eq. (A.21) is done
\begin{equation}
 \sum\limits_{\tilde{\lambda}} \left| M \left( --;
-, \tilde{\lambda} \right)
T \left( \tilde{\lambda} \right)   \right|^2 \to
\frac{1}{2} \sum\limits_{\tilde{\lambda}} \left| M \left( --;
-, \tilde{\lambda} \right) \right|^2 
\sum\limits_{\tilde{\lambda}}\left| T \left( \tilde{\lambda} \right)   \right|^2  
\end{equation}
We can see that spin correlations  do not change the  results substantially 
($ \leq 2$ \% for considered c.m energies and the chargino mass).  
It is important   that the positron angular distributions are
not so strongly peaked in the beam directions, even for $\sqrt{s}=1$ TeV
collider energy. With assumed cuts  $(\vert \cos
\Theta_+ \vert \leq 0.95)$ enough events will be detected to investigate
the process.

The only SM process, which gives a four charged
lepton signal without missing energy is \cite{bac} $e^-e^- \to e^-e^- Z$. 
With a possible Z
boson decay to the lepton-antilepton pair, it does 
not coincide with the process under
investigation ($e^-e^- \to 2 \mu^- e^-e^+$). 
That means that we do not have to bother about the SM background
contamination. However, 
this cross section is large enough  ($\simeq 1$ pb for $ 0.5 \leq \sqrt{s} 
\leq 2$ TeV energies) to cover some other scenarios. 
As an example let us assume that
not only $\lambda_{112}$ but also $\lambda_{121}$ is not negligible
and change the second coupling in the chargino decay channel (Fig.3) from 
$\lambda_{112}$ to $\lambda_{121}$. That means that we have now
$(e^-e^- \to ) e^-e^- \mu^- \mu^+$ in the final state, 
and this scenario will be dominated by the SM
process given above with the Z decay to the muon antimuon pair. 
In this way we can find that $\mu^- \mu^- e^-e^+(\tau^+),
\mu^- \tau^- e^- e^+(\mu^+)$ and $ \tau^- \tau^- e^- e^+$ charged 
lepton signals 
are testable (meaning sensitivity to the products $\lambda_{112} \cdot
 \lambda_{112} (\lambda_{113})$,
$\lambda_{112} \cdot \lambda_{113} (\lambda_{123})$ 
and $\lambda_{113} \cdot \lambda_{113}$, respectively).

Finally,  our results can be easily applied  to the
muon-muon collider where another set of R-parity lepton violating couplings can be
tested, namely: $\lambda_{221} \cdot \lambda_{212}$ 
($e^-e^-\mu^- \mu^+$ in the final
state),
$\lambda_{221} \cdot \lambda_{213}$ ($e^-e^-\mu^-\tau^+$) and
$\lambda_{221} \cdot \lambda_{231}$ ($e^-\mu^-e^+\tau^-$).

\section{Conclusions}

Present experimental limits on  R-parity violating 
couplings do not exclude large and detectable lepton number violating signals
in lepton-lepton collisions.
We discuss such a possibility in conjunction with single chargino production and 
its subsequent leptonic decay. 
If at least one $\lambda$ value is large enough - in our discussion mainly
$\lambda_{112}$ - single chargino production
in electron-electron collisions will be observable  (Fig.2).
If the chargino is  NLSP and sneutrino the LSP, 
a unique lepton number violating signal of
four charged leptons 
without missing energy could  be observed.

\section*{Acknowledgments}

We would like to thank H. Fraas, 
C. Bl\"ochinger, 
S. Hesselbach (W\"urzburg University) and G.~Moortgat--Pick (DESY, Hamburg)
for helpful discussions 
and valuable remarks. 
This work was supported by the Polish Committee for Scientific Research under 
Grant No.~2P03B05418. 
J.G. would like to thank
the Alexander von Humboldt-Stiftung for   fellowship.

\begin{appendix}
\section{Appendix}
The helicity amplitude for 
single chargino production $e^-e^- \to \mu^-(\tau^-) \tilde
\chi^-$ is given by:
\begin{eqnarray}
M(\sigma_1,\sigma_2; \lambda , \tilde{\lambda})&=&
g  \lambda_{112} 
\Biggl[ \bar{u} ({p},\lambda ) (1-\gamma_5) u(k_1,\sigma_1)
\frac{1}{t-m_{\tilde{\nu}_e}^2}  
\bar{u}(\tilde{p},\tilde{\lambda})  (1-\gamma_5) u(k_2,\sigma_2) 
 \nonumber \\
&-&  \bar{u} (\tilde{p},\tilde{\lambda} )  (1-\gamma_5) u(k_1,\sigma_1)
\frac{1}{u-m_{\tilde{\nu}_e}^2}  
\bar{u}({p},\lambda )  (1-\gamma_5) u(k_2,\sigma_2)
\Biggr] , \\
t(u)&=&m_{\tilde \chi}^2-\sqrt{s}(\tilde E \mp 2 \tilde 
p \cos{\tilde \Theta})
\end{eqnarray}
where $(\tilde p, \tilde \lambda)$ denotes the momentum and
helicity of the chargino, $(k_{1(2)}, \sigma_{1(2)})$ are the corresponding
quantities for the incoming electrons, 
$p$  and $\lambda$ denote momentum
and helicity of the muon (tau) and
$( \tilde \Theta, \tilde \phi)$ label the c.m. azimuthal and 
polar angles of a chargino  with respect to the direction of the 
initial electron $e_1^-$.

To work out the helicity amplitudes, we use the
Weyl-- van der Waarden spinor formalism
\cite{van} in which the 4--spinors can be written:

\begin{eqnarray}
u(p,\lambda)&=& \left( \matrix{ \sqrt{E-p \lambda}\; \chi (p,\lambda) \cr 
\sqrt{E+p \lambda} \; \chi (p,\lambda) } \right) , \;\;\;\;
v= \left( \matrix{ \lambda \sqrt{E+p \lambda} \; \chi (p,-\lambda) \cr 
- \lambda \sqrt{E-p \lambda} \; \chi (p,-\lambda) } \right), 
\label{van1}
\end{eqnarray}
and the Weyl spinors are given by:
\begin{eqnarray}
 \chi (p,+1/2)&=& \left( \matrix{ e^{-i\phi/2} \cos{\theta/2} \cr
 e^{i\phi/2} \sin{\theta/2} } \right), \;\;\;\;
 \chi (p,-1/2)= \left( \matrix{ -e^{-i\phi/2} \sin{\theta/2} \cr
 e^{i\phi/2} \cos{\theta/2} } \right),
 \label{van2}
 \end{eqnarray}
where $\Theta,\phi$ denote the azimuthal and the 
polar angle of a particle with respect to the $\hat z$ axis.
 
In the limit of zero mass of all charged leptons
we have only two non-vanishing helicity amplitudes
($\sigma_{1,2},\lambda=-1/2,\tilde{\lambda}=\pm 1/2$), namely:

\begin{eqnarray}
M (- ,- \rightarrow -, - ) &=& g  \sqrt{2s} 
\sqrt{ (\tilde{E}-\tilde{p}) (\tilde{E}-\tilde{p})}
 \lambda_{112}
\left[ \frac{\cos^2{\frac{\tilde{\Theta}}{2}}}
{t-m_{\tilde{\nu}_e}^2}+ 
\frac{\sin^2{\frac{\tilde{\Theta}}{2}}}{u-m_{\tilde{\nu}_e}^2} \right] \\
M (- ,- \rightarrow  - ,+) &=& -g  \sqrt{2s} 
\sqrt{ (\tilde{E}+\tilde{p}) (\tilde{E}-\tilde{p})}
\cos{\frac{\tilde{\Theta}}{2}}\sin{\frac{\tilde{\Theta}}{2}}
\lambda_{112} 
\left[ \frac{1}{t-m_{\tilde{\nu}_e}^2}- 
\frac{1}{u-m_{\tilde{\nu}_e}^2} \right]. \nonumber \\
&& 
\end{eqnarray}

The cross section is:

\begin{equation}
d\sigma =   \frac{1}{2s}  
dLips \left( s, {p},\tilde{p} \right)
\sum\limits_{\tilde{\lambda} } \left| M \left( 
-,-; - ,\tilde{\lambda}_j 
 \right) \right|^2 
\end{equation}

where

\begin{eqnarray}
dLips \left( s, {p},\tilde{p} \right) & =& 
\frac{\tilde p}{16 \pi^2 p} d\cos{\tilde{\Theta}} d \tilde{\phi}.
\end{eqnarray}

In Fig. 3 we study the R-parity violating  chargino decays via
sneutrino exchange in the t-channel:
$\tilde \chi \to e^-e^+ \mu^-$. Analogously to the production
the amplitude is:
\begin{eqnarray}
T(\tilde{\lambda}_i)&=&
 {g} \lambda_{112}
\Biggl[ \bar{u} (p_1,\sigma_1 ) (1-\gamma_5) 
u(\tilde{p},\tilde{\lambda})
\frac{1}{t-m_{{\tilde{\nu}_e}}^2+i \Gamma_{\tilde \nu} m_{{\tilde \nu}_e}}  
\bar{u}(p_2, \sigma_2 ) (1-\gamma_5) v(p_+, \sigma_+) \Biggr], \nonumber \\
\end{eqnarray}
where ($ (p_{1(2)}^i, \sigma_{1(2)}^i)$ denote the  momenta and
helicities of the electron and muon,
$(p_+,\sigma_+)$ are analogous quantities for the final positron.

Using Eqs.~(\ref{van1},\ref{van2})  we get
\begin{eqnarray}
 T(\tilde \lambda = +1/2)&=& \Omega_t(+) \left[
 -e^{i/2(\phi_1-\tilde{\phi})}
 \sin{\frac{\Theta_1}{2}}\cos{\frac{\tilde{\Theta}}{2}}+
 e^{-i/2(\phi_1-\tilde{\phi})}
 \cos{\frac{\Theta_1}{2}}\sin{\frac{\tilde{\Theta}}{2}} \right] \nonumber \\
 & \times & 
 \left[ -e^{i/2(\phi_2-\phi_+)}
 \sin{\frac{\Theta_2}{2}}\cos{\frac{\Theta_+}{2}} +
 e^{-i/2(\phi_2-\phi_+)}
 \cos{\frac{\Theta_2}{2}}\sin{\frac{\Theta_+}{2}} \right]  
 \label{norm1}\\
  T(\tilde \lambda = -1/2)
  &=& \Omega_t(-) \left[ e^{i/2(\phi_1-\tilde{\phi})}
 \sin{\frac{\Theta_1}{2}}\sin{\frac{\tilde{\Theta}}{2}}+
 e^{-i/2(\phi_1-\tilde{\phi})}
 \cos{\frac{\Theta_1}{2}}\cos{\frac{\tilde{\Theta}}{2}} \right] \nonumber \\
 & \times & 
 \left[ -e^{i/2(\phi_2-\phi_+)}
 \sin{\frac{\Theta_2}{2}}\cos{\frac{\Theta_+}{2}} +
 e^{-i/2(\phi_2-\phi_+)}
 \cos{\frac{\Theta_2}{2}}\sin{\frac{\Theta_+}{2}} \right] 
\end{eqnarray}
 where 
$(\Theta_{+(1,2)},\phi_{+(1,2)})$ denote azimuthal and polar angles of
the final positron (electron (1), muon (2)) which are defined with respect to 
the direction of the initial electron beam $e_1$ 
and $\Omega_{t}(\pm)$ is given by
 \begin{equation}
 \Omega_{t}(\pm)= g \sqrt{8 E_+E_1
 E_2} \sqrt{\tilde{E}\pm \tilde{p} \tilde{\lambda}}
\sum\limits_{m_{{\tilde \nu}_n}}\lambda_{n12}
\frac{1}{t-m_{\tilde{\nu}_n}^2+i \Gamma_{\tilde \nu} m_{{\tilde \nu}_n}}. 
 \end{equation}

The decay width can be written as
\begin{equation}
d\Gamma=\frac{1}{2 m_{\tilde \chi}} 
dLips \left( \tilde{m}, p_1,p_2, p_+
\right) \sum\limits_{\tilde \lambda} | T ( \tilde \lambda  )|^2
\end{equation}
where
\begin{eqnarray}
dLips \left( \tilde{m}, p_1,p_2, p_+ \right) &=& 
\frac{1}{\left( 2 \pi
\right)^5 8 } \int dE_+ \int d\cos{\Theta_+} \int d \phi_+ 
\int d\cos{\Theta_2} \int d \phi_2 .
\label{ph}
\end{eqnarray}
Formulae Eq.~(A.13) describe
 the 3--body--decay of a chargino  with energy
$\tilde E$ and angles $\tilde \Theta, \tilde \phi$. The angles of the chargino
and of the particles produced by chargino decay are defined  with respect 
to the 
direction of the initial electron beam $e^-_1$. 
Angles of the decaying particles are 
also defined with respect to the initial CM system of colliding electrons.
We have left quantities connected with $e^+$ as independent parameters to 
be integrated over. 
From the 12 quantities describing the chargino 3--body--decay, 
four are eliminated by  momentum
conservation. These are chosen to be the angles
$\Theta_1, \phi_1$ and the energies $E_{1,2}$, namely:
\begin{eqnarray}
E_2&=& \frac{1}{2} 
\frac{\tilde{m}_i-2 \tilde{E} E_+ +2 \tilde{p} E_+ 
\cos{ \left( \tilde{p},p_+ \right) }}{ \tilde{E}-E_+- \tilde{p}
 \cos{ \left( \tilde{p},p_2 \right) } + 2 E_+
 \cos{ \left( {p}_2,p_+ \right) }} \\
 E_1&=&\tilde{E}-E_2-E_+ \\
 \cos{\Theta_1}&=&\frac{\tilde p \cos{\tilde \Theta }-E_2 \cos{\Theta_2}
-E_+ \cos{\Theta_+}}{E_1}.
\end{eqnarray}
The angle $\phi_1$ is fixed by two relations:
\begin{eqnarray}
E_1 \sin{\Theta_1} \cos{\phi_1}&=&\tilde p 
\sin{\tilde \Theta } \cos{\tilde \phi}-
E_2 \sin{\Theta_2} \cos{\phi_2}
-E_+ \sin{\Theta_+} \cos{\phi_+}, \\
E_1 \sin{\Theta_1} \sin{\phi_1}&=&\tilde p 
\sin{\tilde \Theta } \sin{\tilde \phi}-
E_2 \sin{\Theta_2} \sin{\phi_2}
-E_+ \sin{\Theta_+} \sin{\phi_+}
\end{eqnarray}

We end up with the 8 parameters
(these are given by Eq.~(\ref{ph}) and $\tilde E,\tilde \Theta, 
\tilde \phi$).

For completeness, it is trivial to compute the sneutrino decay width. For
 $m_{\tilde \nu} \leq m_{\tilde \chi}$ (the scenario discussed in the text,
see Fig.4)  only one decay channel
to an $e^- \mu^+$  pair is open (for simplicity we assume that only
one $\lambda_{112}$ dominates):
\begin{equation}
\Gamma_{\tilde \nu}=\lambda_{112}^2 m_{\tilde \nu_e}/8 \pi .
\end{equation}

Finally for the combined process of production and decay we obtain in the 
narrow width approximation:
\begin{eqnarray}
d\sigma (e^-e^- \rightarrow 2 \mu^- e^-e^+)&=&  
\frac{1}{2s}dLips \left( s, {p},\tilde{p} \right)
dLips \left( \tilde{m}, p_1,p_2, p_+ \right)  \frac{1}{2 \pi}
\nonumber \\
&& \sum\limits_{\tilde{\lambda}} \left| M \left( --;
-, \tilde{\lambda} \right)
T \left( \tilde{\lambda} \right)   \right|^2 
\left( \frac{ \pi}{m \Gamma} \right).
\label{full}
\end{eqnarray}
\end{appendix}

\end{document}